\documentclass[iop]{emulateapj}
\usepackage{amsmath}
\usepackage{apjfonts}
\usepackage{amssymb}
\usepackage{xspace}
\usepackage{hyperref}
\usepackage{gensymb}
\usepackage{xstring}
\usepackage{natbib}
\usepackage{multirow}
\usepackage{graphicx}
\usepackage{epstopdf}
\usepackage{epsf}
\usepackage{subfigure}
\usepackage[normalem]{ulem}
\usepackage[usenames,dvipsnames]{xcolor}
\usepackage[utf8]{inputenc}

\def\al{\alpha}
\def\z6{z\sim6 }
\def\fon{f_{\rm qso}}
\def\fact{f_{\rm on}}
\def\fduty{f_{\rm duty}}
\def\Msun{\rm M_\odot}
\def\Vmax{V_{\rm max}}
\def\Vcirc{V_{\rm circ}}
\def\rmax{r_{\rm max}}

\def\imax{i_{\rm max}}
\def\M1450{M_{\rm 1450}}

\begin{document}
\title{Constraints on Duty Cycles of Quasars at $\z6$}
\author{Huanqing Chen\altaffilmark{1,2}, Nickolay Y.\ Gnedin\altaffilmark{3,1,2}}

\altaffiltext{1}{Department of Astronomy \& Astrophysics, The University of Chicago, Chicago, IL 60637, USA; hqchen@uchicago.edu}
\altaffiltext{2}{Kavli Institute for Cosmological Physics, The University of Chicago, Chicago, IL 60637 USA;}
\altaffiltext{3}{Particle Astrophysics Center, Fermi National Accelerator Laboratory, Batavia, IL 60510, USA}

\begin{abstract}
We study the mass of quasar-hosting dark matter halos at $\z6$ and further constrain the fraction of dark matter halos hosting an active quasar $\fact$ and the quasar beaming angle $\imax$ using observations of CII lines in the literature.
We make assumptions that (1) more massive halos host brighter quasars, (2) a fraction of the halos host active quasars with a certain beaming angle, (3) cold gas in galaxies has rotational velocity $\Vcirc=\alpha \Vmax$, and that (4) quasars point randomly on the sky. We find that for a choice of specific $\alpha \gtrsim 1$, the most likely solution has $\fact < 0.01$, corresponding to a small duty cycle of quasar activity. However, if we marginalize over $\alpha$, for some choices of a prior a second solution with $\fact=1$ appears. Overall, our the constraints are highly sensitive to $\alpha$ and hence inconclusive. Stronger constraints on $\fact$ can be made if we better understand the dynamics of cold gas in these galaxies.
\end{abstract}
\keywords{reionization, quasars: general}

\section{Introduction}
 
Despite the mounting evidence that the Epoch of Reionization (EoR) ends at $\z6$, many details on how the universe gets reionized remain poorly understood. One of the outstanding problem in the EoR research is what contribution of quasars is to the total ionization history. Although many studies suggest that quasar may not play a major role in driving reionization on large scales \citep{onoue17}, quasars dominate the reionization process in their local environments \citep{lidz07,kakiichi17}.
Their high ionizing luminosities create proximity zones where H and He are predominantly ionized. In their proximity zones, galaxy formation may be delayed \citep{kitayama00}. How many galaxies are impacted by quasars is closely related to quasar life time, opening angles, and more importantly how over dense the quasar environment is, which unfortunately are all very uncertain.

Due to the increased halo clustering near higher mass halos,  
the mass of a quasar-hosting halo determines the number of nearby galaxies that may be impacted by the quasar. Therefore, to study reionization history in quasar surroundings, it is important to understand how massive their host halos are.

Masses of quasar-hosting halos can be estimated in many different ways \citep{martini04}. In the local universe, one can connect the black hole mass to the host halo mass by using the tight M-$\sigma$ relation \citep{ferrarese00,gebhardt00} and the correlation between the circular velocity and the bulge velocity dispersion \citep{ferrarese02}. 
At higher redshifts, one can measure the relation between the quasar luminosity and the clustering length, which in turn can be translated into the mass of quasar-hosting halos \citep{martini01,haiman01}. For example, \citet{shen07} used SDSS data and found a minimum halo mass $(4-6)\times10^{12}$ $\Msun$ at $3.5<z<5.3$ for their sample.

However, at z$\sim 6$ studies of quasar host halos are rare. The number of quasars detected is so limited that the clustering method can not be applied yet. 
Fortunately, recently ALMA opened a new window to such studies.
Thanks to its exceptionally high spatial and spectroscopic resolution, it is possible to map many $\z6$ quasar host galaxies in millimeter molecular lines and measure their emission line profiles. Emission lines from cold gas offer dynamical information, which is crucial in determining the halo mass.

Measuring quasar halo masses has further fascinating application of constraining quasar duty cycles. If we assume all dark matter (DM) halos host a quasar, then small quasar-hosting halo mass indicates a large number of quasars are not active, suggesting a small quasar duty cycle. Using clustering method, \citet{shen07,shankar10} found increasing duty cycle from $\fduty \sim 0.01$ at $z=1.45$ to $0.03-0.6$ at $3.5<z<5.3$. 

In this paper, we use updated observational data from the literature (mainly ALMA observations) to estimate quasar-hosting DM halo mass. We use a simple model to put constraints on the fraction of DM halos that host a visible quasar $\fon$ and the quasar opening angle $\imax$.  We describe our methods in Section \ref{sec:method} and show the results of constraining $\fon$ and $\imax$ in Section \ref{sec:result}. Discussion and conclusions are presented in Section \ref{sec:discuss} and Section \ref{sec:conclusion}.

\section{Methodology}\label{sec:method}

Our main goal is to study the relation between the quasar luminosities and the mass of DM halos they reside in at $\z6$, and further constrain the quasar duty cycle and the opening angle. For this purpose, we make several assumptions described below, and use CII line profiles of 44 known $\z6$ quasars in \citet{decarli18} to estimate the halo mass. 

\subsection{Assumptions}
We list our assumptions hereafter and return to 3 and 4 in the Discussion: 
\begin{enumerate}
\item Every DM halo hosts a SMBH, and more massive halos host brighter quasars.
\item Among all DM halos, a fraction of them ($\fact$) host active quasars, and quasar emission is beamed with a solid angle of $\Omega$.
\item Cold gas in quasar-hosting galaxies is rotationally supported with a flat rotation curve with $\Vcirc$.
\item The distribution of quasar orientation is random with a maximum inclination angle $\imax$; $\imax$ and $\Omega$ are related by:
$$\Omega=(1-\cos(\imax))\times4\pi.$$
\end{enumerate}

Our key assumption is that brighter quasars reside in more massive halos. This is probable because large halos at $\z6$ must experience fast accretion, which may also indicate fast quasar fueling process \citep{martini01}. Observations also support this assumption and indicate a narrow scatter \citep{shankar10}. We adopt an abundance-matching ansatz:
\begin{equation}
n(>L)=\fon n(>M_h),
\end{equation}
where $n(>L)$ is the number density of quasars with luminosity larger than $L$ and $n(>M_h)$ is the number density of DM halos with mass larger than $M_h$.
The parameter $\fon$ is the fraction of halos that host visible quasars. Due to the fact that quasars have limited lifetime and are obscured when inclination angles are large, $\fon$  should be less than one. In an idealized model, $\fon$ is simply
\begin{equation}{\label{fml:fon}}
\fon = \fact\frac{\Omega}{4\pi}.
\end{equation}

By matching the number densities of quasars with the number density of DM halos, we can obtain the relationship between quasar luminosity and the underlying halo DM mass:

\begin{equation}
\int_{L}^\infty \phi(L) \ dL =\fon \int_{M_{h}}^\infty \phi(M_h)\  dM_h.
\label{eq:am}
\end{equation}

We use $\phi( L)$ from \citep{onoue17}. We note that although the faint end of the quasar luminosity function is still uncertain, this does not affect our results because the observed quasars mainly populate the bright end.

\subsection{Halo Mass Function}

There exist a number of analytical fits for halo mass functions. Unfortunately, analytical fitting formulas at $z\sim6$ are of considerable differences \citep{lukic07,tinker08,reed13,watson13,murray13}. For this reason, we use the halo catalog of our existing CROC simulations \citep{gnedin14} to determine what analytical halo mass function to use. We find that Tinker08 \citep{tinker08} halo mass function agrees exceptionally well with numerical simulations of $\z6$ universe in the mass range of $10^9 - 10^{12} \Msun$ (the massive halo regime), both in terms of amplitude and shape.

\begin{figure}[!htb]
  \includegraphics[width=1\linewidth]{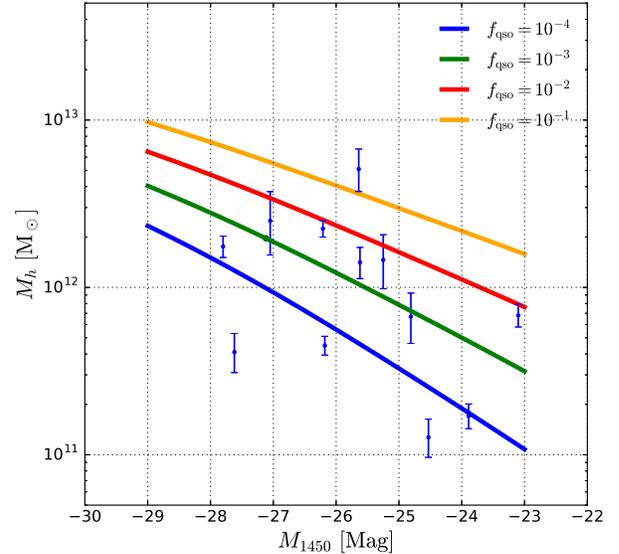}\hspace{-25pt}

\caption{Abundance matching results for different values of the parameter $\fon$. Smaller $\fon$ results in smaller halo mass. Blue dots are quasars in literature with CII maps available and galactic disks marginally resolved. Error bars only account for the uncertainties in FWHM measurements.}\label{fig:am}
\end{figure}

Using the Tinker08 halo mass function and the quasar luminosity function from \citet{onoue17}, we obtain the function of quasar luminosity and halo mass with different parameter $\fon$, shown in Figure \ref{fig:am}.
Smaller $\fon$ results in smaller halo mass because more halos are required to match the observed quasar number density at certain $\M1450$.

\subsection{Estimating the Virial Mass}
 
To obtain the virial mass from observations requires several assumptions. Observations typically only provide some velocity information from spectra. We need to find a relation between mass and spectroscopic observables.

In DM-only simulations, the maximum of circular velocity $\Vmax$ is tightly correlated with the virial mass of DM halos. Using the publicly available 250 ${\rm Mpc}/h$ Bolshoi halo catalog \citep{klypin11,behroozi13a,behroozi13b}, we fit the $\z6$ halos and found a best-fit power-law relation of
\begin{equation}{\label{fml:vmaxmvir}}
\log_{10} \frac{\Vmax}{\rm km/s}=0.338 \log_{10} \frac{M_{\rm vir}}{\Msun} - 1.55
\end{equation}
with a scatter of 4\% for halos above $10^{11} \Msun$. 
 
We denote $\rmax$ as the radius at which $\Vmax$ is reached. For a  $1\times10^{12}$ $\Msun$ halo at $\z6$, the virial radius is $\sim 70$ pkpc, and $\rmax$ is $\sim 30\%$ of the virial radius. 

In order to connect gas motions to $\Vmax$, we assume that gas has a flat circular velocity profile with $\Vcirc = \al \Vmax$ inside $\rmax$. We treat $\al$ as one of the key parameters and deal with it in Sec. 3.

\begin{figure*}[!htb]
\includegraphics[width=0.5\linewidth]{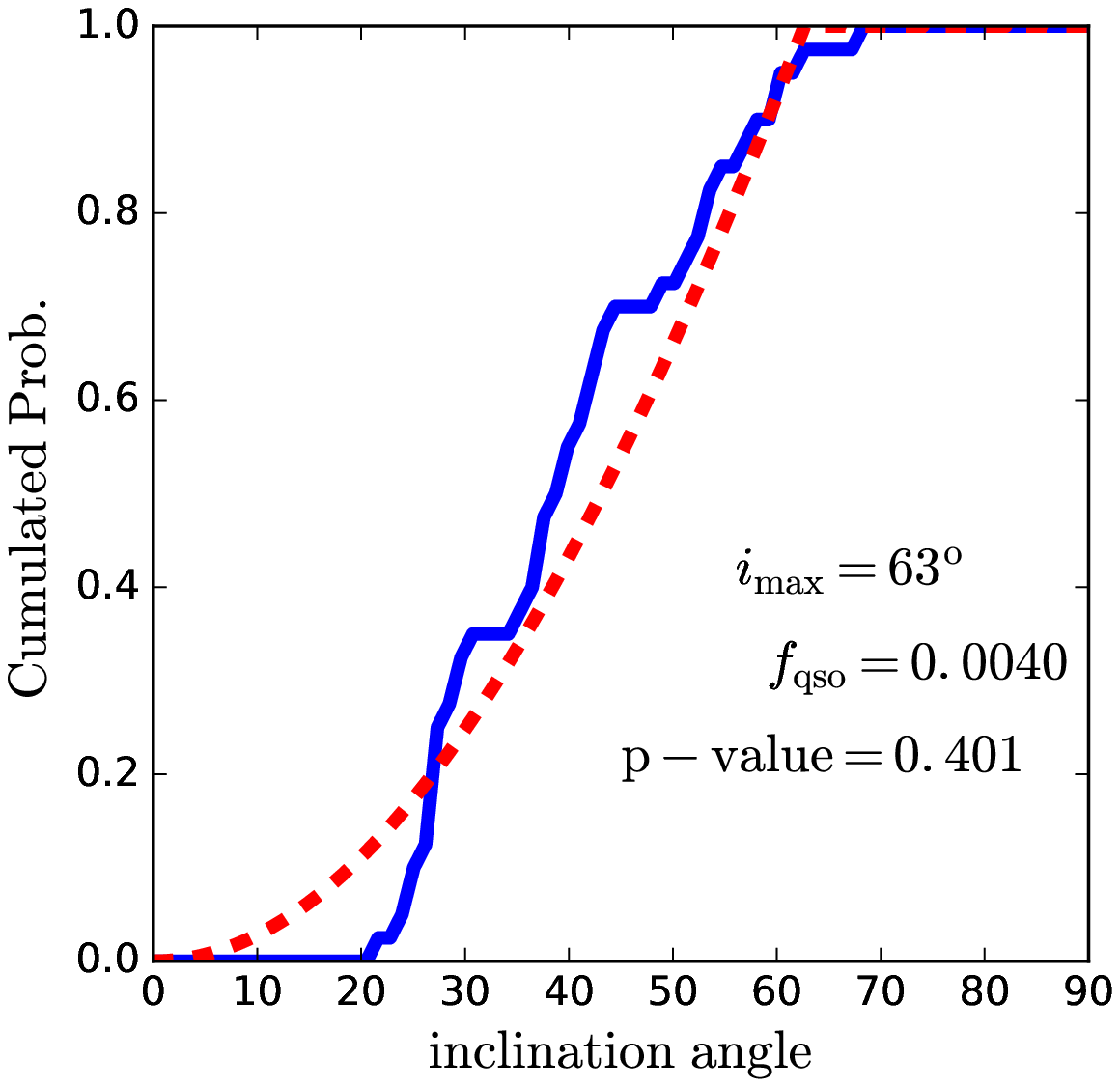}
\includegraphics[width=0.5\linewidth]{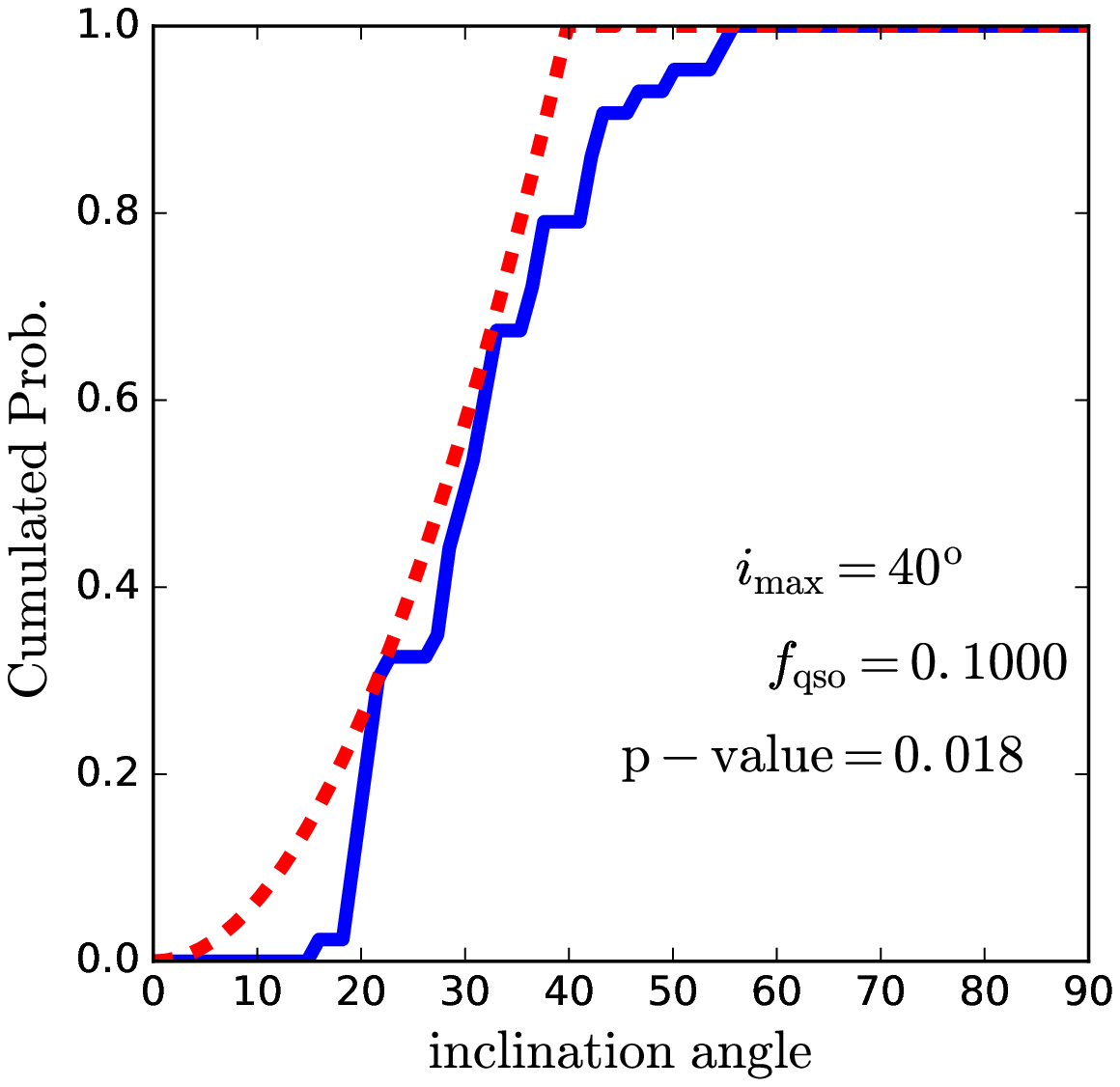}
\caption{Comparison of the derived and expected cumulative inclination angle distributions for two representative sets of values $(\fon,\imax)$=(0.004,63$\degree$) and (0.1,40$\degree$) respectively.}\label{fig:imax}
\end{figure*}

\subsection{Compilation of Observation Data}
Singly ionized carbon CII line at 158 $\mu m$ traces cold gas, which is thought to be bound and rotationally supported within the dark matter halo \citep{wang13}. For $\z6$ galaxies, this line redshifts to mm-wavelength, which can be observed by radio interferometers like ALMA. We thus use CII spectra to constrain halo masses. 

We use the list of 44 $\z6$ quasars in \citet{decarli18} Table 2 and 4, particularly the FWHM of CII emitted from their host galaxies.

Viewing angle affects the line width.  In practice, inclination angles (the angles measured between the line of sight and the normal vector of the disk) are usually estimated by the semi-minor to semi-major axis ratio $i=arccos(b/a)$. If we use the inclination angle measured this way, and using \citep{decarli18,wang13,willott15}
\begin{equation}
\Vmax=\Vcirc=0.75 {\rm FWHM}/\sin i,
\end{equation}
we can estimate the halo mass. 
In the literature, there are 12 quasars with marginally resolved disks \citep{willott15,venemans16,wang13,willott13,stefan15,willott17,venemans17,decarli18}. We calculate their halo masses and overlay them as blue points on Figure \ref{fig:am}, with error bars including only the observational errors on the FWHM. 
Most data lie around $\fon=0.001-0.01$ lines, albeit the spread is wide.

However, limited by current observations resolution, the length of galaxy axises can at best be marginally measured at $\z6$: most of the 12 galaxies above have spatial sizes only less than two beam sizes of the telescope.
The rest of quasar-hosting galaxies in the sample (32 out of 44) are not resolved at all. 
Considering the large uncertainty in the inclination angle measurements, we use another method to put constraints on $\fon$ using all 44 galaxies in the sample. 
We assume that
the orientation of quasar beams are random on the sky, with a maximum inclination angle $\imax$. Using a threshold $\imax$ is motivated by the limited opening angle of quasar, usually around $45\degree$ \citep{zhuang18}. 

The cumulative probability distribution of the inclination angle for random orientation is the area of a sphere where the azimuthal angle is smaller than $i$, divided by the total area of the sphere with the maximum azimuthal angle $\imax$:
\begin{equation}{\label{fml:cdf}}
{\rm CDF}=\frac{\int_0^i \sin(i)di}{\int_0^{\imax} \sin(i)di}=\frac{1-\cos(i)}{1-\cos(\imax)}
\end{equation}

For a fixed parameter $\fon$, for each observed quasar we can calculate the inclination angle $i$ such that the quasar lies on the relation set by the abundance matching ansatz (\ref{eq:am}). To constrain $\fon$, we use K-S test to compare the distribution of so obtained $i$ to the assumed random distribution with maximum inclination angle $\imax$. In the following subsection, we show the results of the K-S test. 

We show examples of this distribution in Figure \ref{fig:imax} for two parameters combinations $(\fon,\imax)=(0.004, \ 63\degree)$ and $(0.1,\ 40\degree)$. In the left panel, the calculated cumulative distribution function (CDF) of $i$ (blue solid line) lies close to the model distribution (red dashed line), with a p-value of 0.4, suggesting that the assumption that quasars are randomly oriented is highly probable, while in the right panel the calculated CDF lies systematically below the red model CDF, with a small p-value of 0.018. This implies that either $\imax$ is underestimated or $\fon$ is overestimated.

\section{Constraints on the Visible Quasar Fraction}\label{sec:result}

\begin{figure*}[t]
\includegraphics[width=0.34\linewidth]{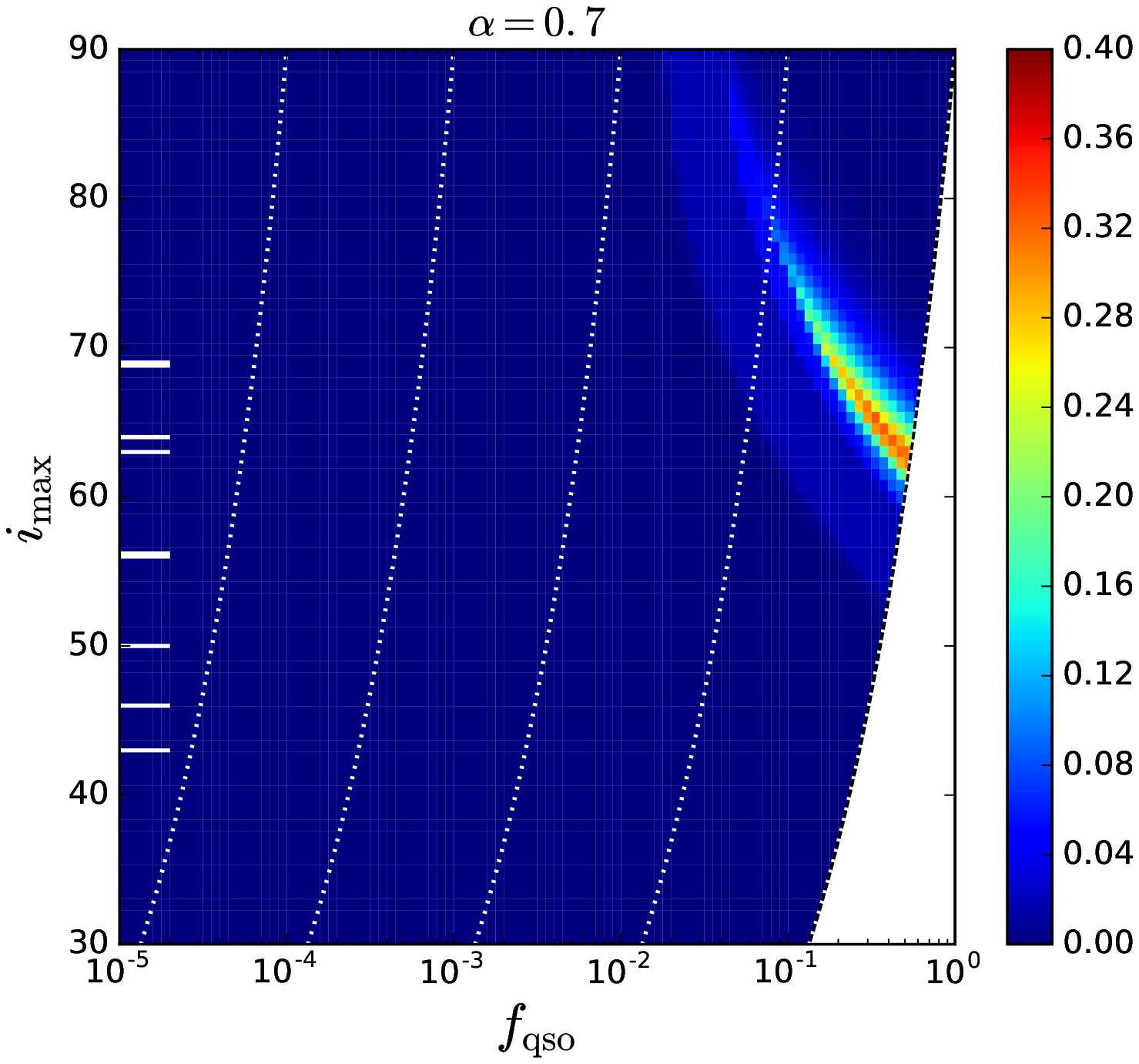}
\includegraphics[width=0.34\linewidth]{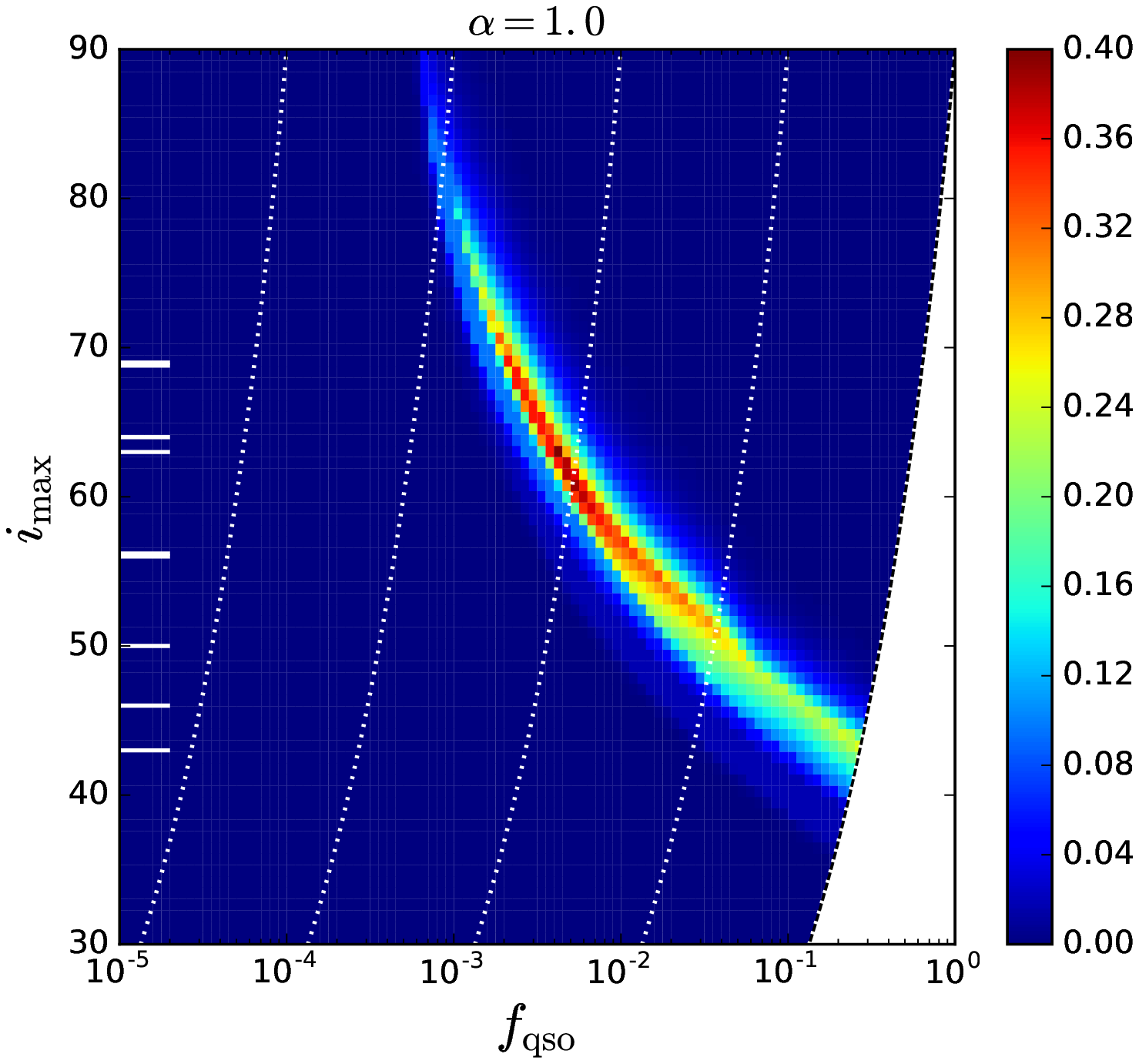}
\includegraphics[width=0.34\linewidth]{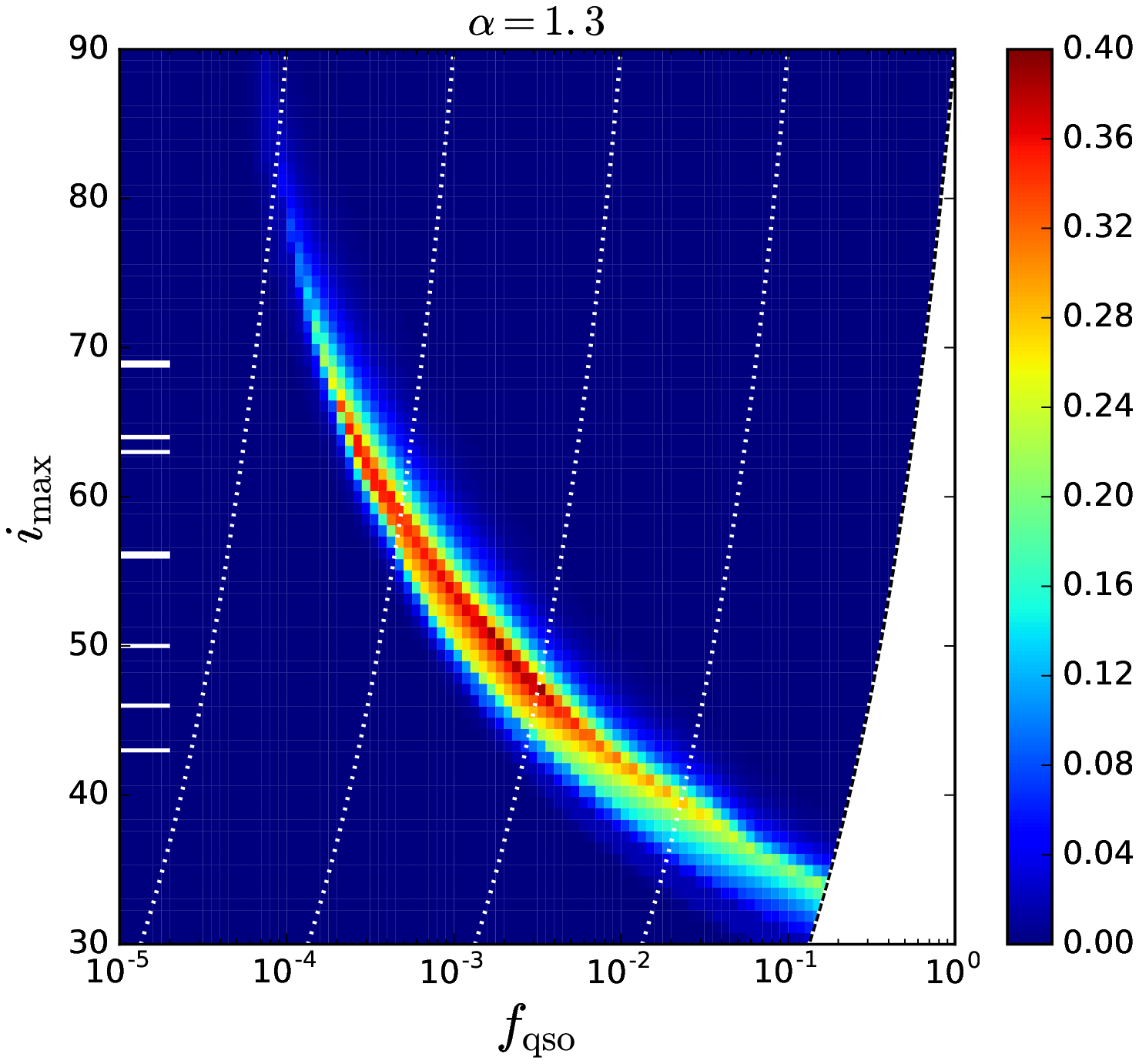}
\caption{P-values of K-S tests for different combinations of $(\fon,\imax)$ for three particular choices of $\alpha\equiv \Vcirc/\Vmax = 0.7$, $1.0$, and $1.3$ (left to right). Note that $\fon$ and $\imax$ are strongly correlated, because larger $\fon$ can be counteracted by a smaller quasar opening angle. Dotted lines show different values of $\fact$ (\ref{fml:fon}): from left to right the values are $\fact=10^{-4},10^{-3},10^{-2},10^{-1}$ respectively. The lower right region is prohibited due to the constraint $\fact\leq1$ (Equation \ref{fml:fon}). Solid white horizontal lines mark observed inclination angles from resolved ALMA galaxies. 
}\label{fig:fonimax}
\end{figure*}

\begin{figure*}[t]
\includegraphics[width=0.34\linewidth]{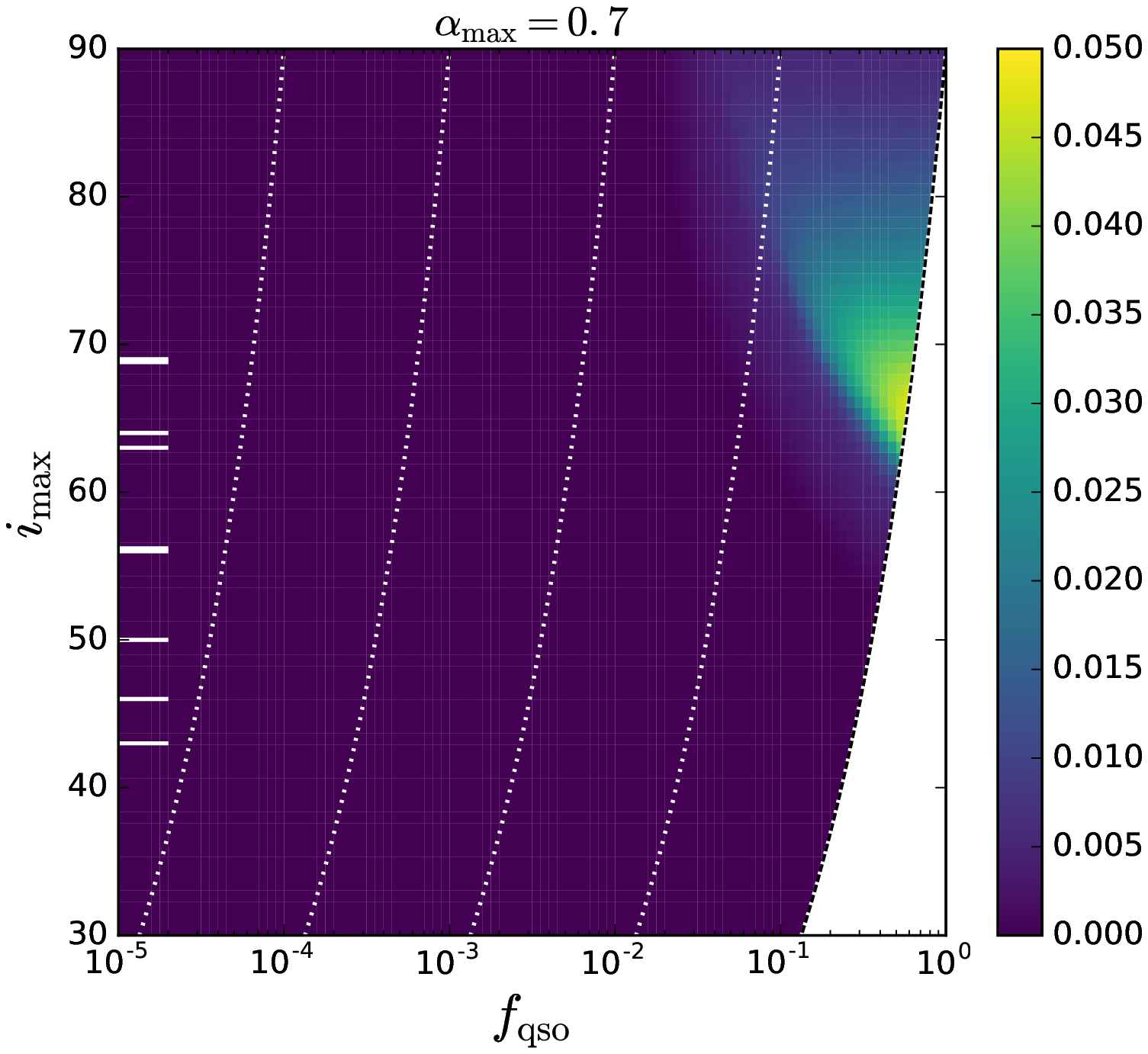}
\includegraphics[width=0.34\linewidth]{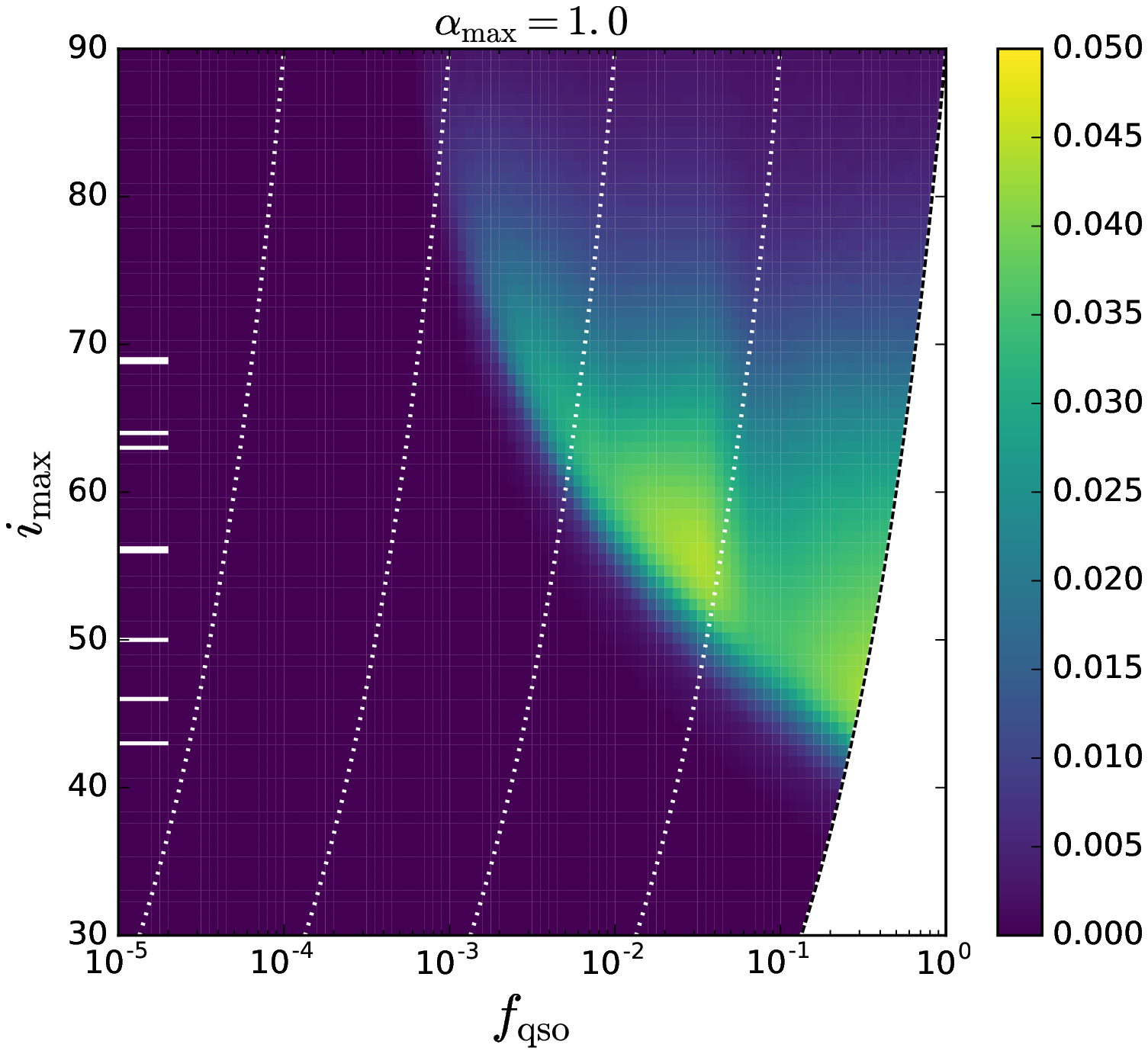}
\includegraphics[width=0.34\linewidth]{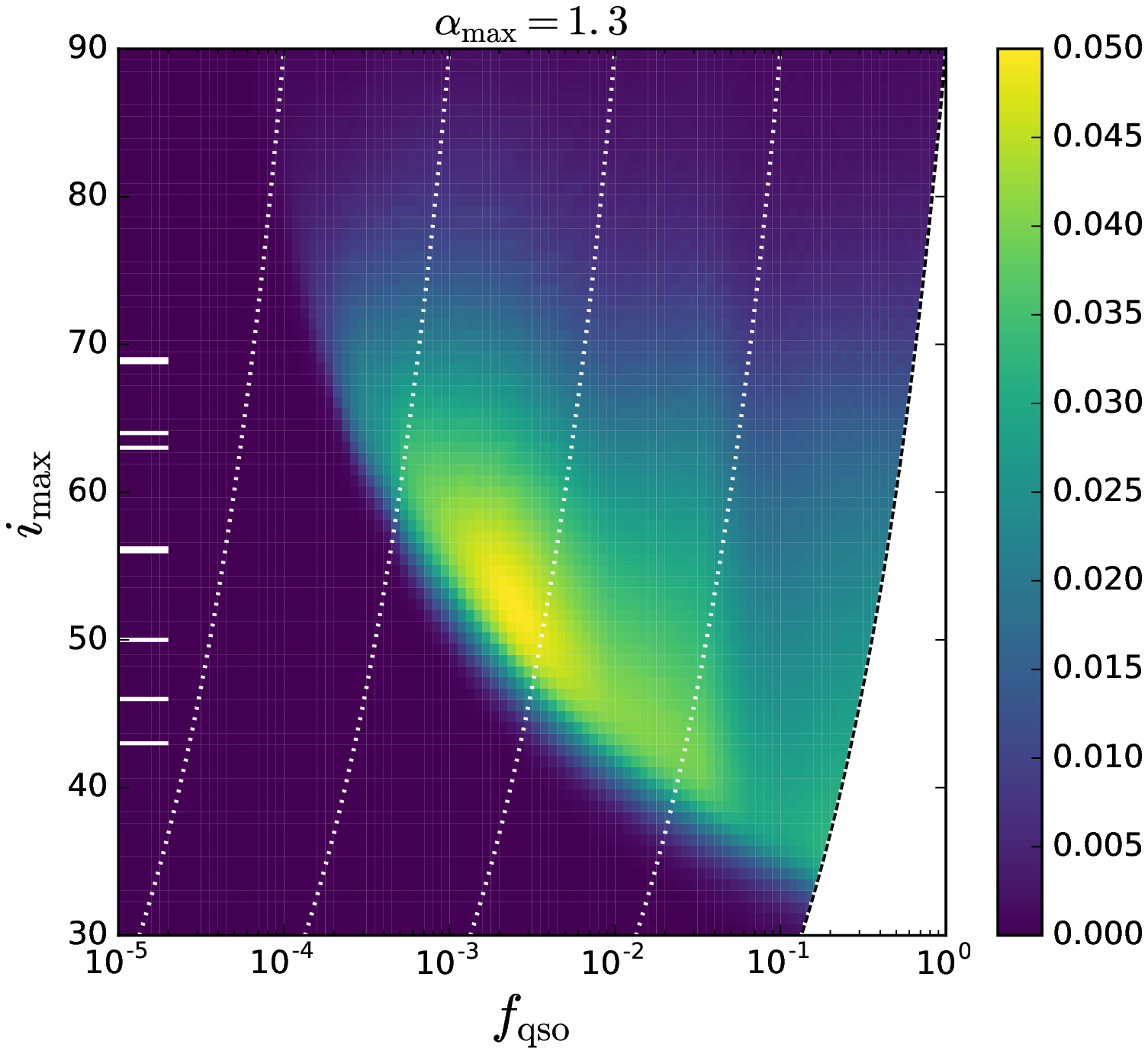}
\caption{Same as Fig.\ \ref{fig:fonimax}, but now with p-values marginalized over $\alpha$. 
}\label{fig:marg}
\end{figure*}

In Figure \ref{fig:fonimax} we show the p-values for different combinations of $\fon$ and $\imax$ for three choices of $\alpha\equiv \Vcirc/\Vmax$. The value of $\alpha=1$ can be considered as a characteristic one, but it can be smaller if the CII emitting gas occupies only the very inner part of a halo, and does not dominate the gravity within $\rmax$. It can also be larger than 1, as low redshift large spiral galaxies often have $\alpha=1.2-1.4$ \citep{nfw97}. We note that there is no acceptable solution for $\alpha<0.6$.

As can be seen from the figure, $\fon$ and $\imax$  are strongly correlated, because larger $\fon$ corresponds to larger halo masses for fixed $\M1450$; in order to obtain the observed FWHM, the assumed inclination angles must decrease, and in order to match the observed $i$ distribution, $\imax$ must decrease as well. For $\alpha\geq1$ the maximum p-value of occurs at small values of $\fon$, suggesting that quasars at $\z6$ reside in fairly low mass halos (See Figure \ref{fig:am}).

The short white lines on the left side of Figure \ref{fig:fonimax} are values of the inclination angle $i_{\rm img}$ estimated from the CII maps of marginally resolved quasar host galaxies (thicker lines represent two galaxies that have the same measured inclination angle). The maximum opening angle $\imax$ should lie close to the few largest $i_{img}$. 

The most probable values of $\fon$ and $\imax$ depend strongly on the choice of $\alpha$. One possible way to remove that dependence and to account for the lack of knowledge in the proper value of $\alpha$ is to "marginalize" over it,
\[
  p_{\rm KS, marg}(\fon,\imax) = \int_0^{\alpha_{\rm max}} w_\alpha p_{\rm KS}(\fon,\imax|\alpha) d\alpha ,
\]
where $w_\alpha$ is the prior on $\alpha$. Notice, that it is namely $p_{\rm KS}$ that should be integrated, as it is a measure of the probability density and a direct analog of a likelihood in this case. In that case, however, we need to adopt an upper limit to the integral, $\alpha_{\rm max}$.

The so marginalized distributions of p-values are shown in Figure \ref{fig:marg} for the three choices of $\alpha_{\rm max}$. In this case, in addition to a low $\fact$ solution, a solution with $\fact$ also appears for $\alpha_{\rm max} = 1$, which becomes the sole one at lower $\alpha_{\rm max}$.

\section{Discussion}\label{sec:discuss}
\subsection{Review of the Assumptions}
\begin{figure}[!htb]
  \includegraphics[width=1\linewidth]{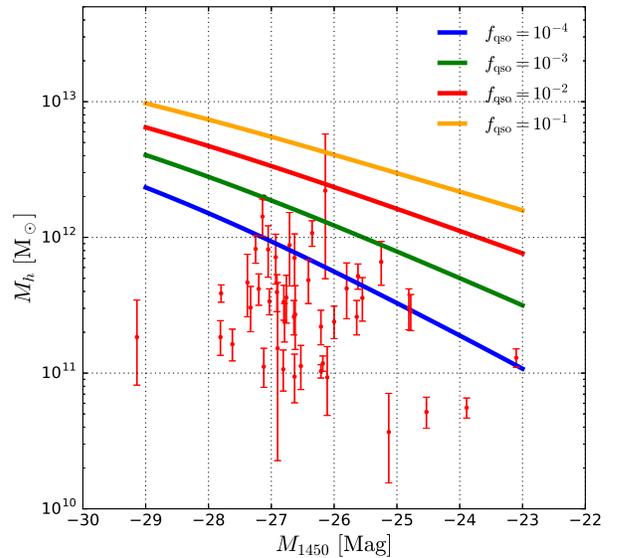}\hspace{-25pt}

\caption{Same as Figure \ref{fig:am}, but with red dots showing quasar masses calculated under the assumption of random gas motion. Error bars are purely from uncertainty in the FWHM measurements. The data spread widely in y-direction and does not show any trend.}\label{fig:amrandommotion}
\end{figure}
To obtain Figure \ref{fig:am} we have made several assumptions. 
Here we discuss assumption 3 and 4 and check how results change when we vary these assumptions.

\subsubsection{Random vs ordered motions}

The assumption that cold gas motion is rotationally supported is motivated by the clear velocity gradients found in three galaxies in \citet{wang13}. Simulations of high redshift galaxies also clearly show that gas forms thin disks \citep{rosdahl18}.

However, if we assume otherwise that CII emitting gas has totally random motions, then the line profile is independent of the viewing angle. Hence, the halo mass can be estimated by:
\begin{equation}{\label{fml:fwhmvmax}}
{\rm FWHM}=2.355\sigma=1.665 \Vcirc
\end{equation}
We overlay the mass calculated this way as red points in Figure \ref{fig:am}, with halo mass uncertainties estimated from the FWHM measurement errors only. The observational points spread widely in y-direction and do show any significant trend. 
This implies that the random gas motion assumption is not valid.

\subsubsection{Random orientation model of quasar}

Random orientation of quasar should be a good assumption, since there is no reason to expect quasar axes to be aligned. However, since different quasars may have different opening angles, the CDF may differ from Equation \ref{fml:cdf} near $\imax$. A more complicated model may use some tapering functions near $\imax$, but the existing data could not constrain it well.

Tapering near $\imax$ may change the K-S test results and impact the most likely value of $\fon$ and $\imax$. To examine the impact of such a modification to our approach, we use a simple linear function between (0.9 $\imax$,CDF(0.9$\imax$)) and (90$\degree$, 1). Using this model, the most likely $\fon$ and $\imax $changed from (0.004, 63) to (0.002,68) and $\fact$ is still $<1\%$.

\section{Conclusions}\label{sec:conclusion}

Our attempt to constraint the high redshift quasar lifetimes using the observed values for the velocities of CII emitting gas is not entirely conclusive. We find that the final answer depends extremely strongly on the assumption of how CII emitting gas populate galactic halos. For the choice of $\al\equiv\Vcirc/\Vmax=1$, our results shows a very small value of $\fact=0.007$, which decreases further if $\alpha$ is allowed to exceed unity. When we marginalize over $\alpha$ with a flat prior, a second solution appears with $\fact=1$; that solution, however, disappears if the adopted range for allowed $\alpha$ extends to values above 1. If each DM halo does indeed host a quasar and the fraction of active quasar at $\z6$ reflects the time fraction spend in quasar phase for each DM halo $\fact=\fduty$, the low $\fact$ value implies a surprisingly small duty cycle at $\z6$. Considering that the mass of a supermassive black hole at $\z6$ can be more massive than $10^9 \Msun$ \citep{wu15}, the small duty cycle implies that the accretion e-folding time must be very small. That further implies that the large fraction of SMBH accrete mass with no significant emission. Another explanation for this small $\fact$ is that only a small portion $f_{\rm BH}$ of halos host a SMBH: $\fact=f_{\rm BH}f_{\rm on|BH}$. 

The small value of $\fact \lesssim 0.01$ is still  consistent with the value of quasar lifetime of of $10^7$ yr that has been found in studies using the transverse proximity effect \citep{borisova16}. Using quasar clustering, \citet{shen07,white08} found a much larger duty cycle approaching unity. All these studies, unfortunately, are subjects to significant uncertainties. For example, the clustering method suffers from the large measurement uncertainties, while our constraint on $\fact$ is strongly sensitive to the assumption of how CII emitting gas populate its host halos. 

The second solution with $\fact=1$ may be more "appealing", as it is consistent with theoretical expectations that high redshift quasars must maintain high accretion rates to reach the observed values for their central SMBH by $\z6$, and is in better agreement with the clustering value. However, our main conclusion is that \emph{at present the data do not select one or the other solution}. It is also possible that both solution are realized in nature as two separate modes of quasar activity.

A significant improvement on this study is possible, if one can constrain the distribution of $\alpha$, for example from high resolution numerical simulations. A purely observational improvement is possible if more of ALMA detected galaxies become resolved with reliable determinations of their inclination angles.

\acknowledgements

Fermilab is operated by Fermi Research Alliance, LLC, under Contract No. DE-AC02-07CH11359 with the United States Department of Energy. This work was partly supported by a NASA ATP grant NNX17AK65G. Some of the simulations and analyses used this paper have been carried out using the Midway cluster at the University of Chicago Research Computing Center, which we acknowledge for support. This research also used resources of the Argonne Leadership Computing Facility, which is a DOE Office of Science User Facility supported under Contract DE-AC02-06CH11357. An award of computer time was provided by the Innovative and Novel Computational Impact on Theory and Experiment (INCITE) program. This research is also part of the Blue Waters sustained-petascale computing project, which is supported by the National Science Foundation (awards OCI-0725070 and ACI-1238993) and the state of Illinois. Blue Waters is a joint effort of the University of Illinois at Urbana-Champaign and its National Center for Supercomputing Applications. This work made extensive use of the NASA Astrophysics Data System and {\tt arXiv.org} preprint server.

\bibliography{ms}

\end{document}